# Field driven Metal-Insulator transition in rhombohedral Bismuth and Arsenic crystals

N. K. Karn[1,2,*], Mukul S. Laad[3], and V.P.S. Awana[1,2]

[1]*CSIR-National Physical Laboratory, Dr. K. S. Krishnan Marg, New Delhi-110012, India*
[2]*Academy of Scientific and Innovative Research (AcSIR), Ghaziabad 201002, India*
[3]*Institute of Mathematical Sciences, Taramani, Chennai 600113, India*

**Abstract**

The metal to insulator (MIT) transition is accompanied by huge changes in physical responses by the control and tuning of experimental parameters like doping, pressure, chemical composition, and magnetic field. Here, we study the magnetic field-driven MIT for two pnictides in their elemental form, namely Arsenic and Bismuth. At low temperatures, Bismuth shows an unusual behaviour of a re-entrant IMT at high fields in addition to a higher temperature MIT at smaller fields. However, Arsenic shows the commonly observed single MIT. The Shubnikov de Haas (SdH) oscillations are observed for both As and Bi below 10 K. Giant magneto-resistance of the order of ~$10^5$ (MR%) is observed for both crystals at 2 K and 14 Tesla transverse magnetic field. The unusual Kohler scaling behaviour of MR at low temperature indicate the presence of increased carrier density attributed to the melting of excitons. Based on a microscopic model, the microscopic processes underpinning the unusual features of a field-driven MIT and re-entrant IMT, along with the relevance of both excitonic and Bose metal correlations near these incipient instabilities, are qualitatively described in the framework of field-driven excitonic condensate and Das-Doniach preformed pair scenarios in one single picture.

***Corresponding Author**
Navneet Kumar Karn
Senior Research Fellow, CSIR
CSIR-National Physical Laboratory,
Dr. K.S. Krishnan Marg, 110012, India
E-mail: nkk15ms097@gmail.com
Ph. +91-11-45609357

## Introduction

The metal to insulator (MIT) phase transition is a fascinating phenomenon marked by transition from highly conductive metallic state to near insulating resistive state. Such a transition can be driven by temperature (T) [1], pressure (P) [2, 3] or by doping (x) [4, 5] following various mechanisms viz. band-gap induction by strong correlation effects [6, 7], leading to a Mott insulator. Detailed theoretical and experimental literature on the MIT can be found in reviews by Mott [6] and Imada [7]. In the last decade, MIT driven by transverse magnetic field has been observed in several novel semi metallic transition metal dichalcogenides $MX_2$ (M = Transition metal, X = chalcogen) [8, 9], TPn (T = Nb and Ta, Pn = P and As) [10, 11], ReBi/Sb(Re=Rare earth) [12-14], and even in the pnictide elements [15-18]. Among these materials, some are magnetic, while others are non-magnetic, but they seemingly show a similar type of MIT [8-18]. At even lower T, a plateau region (metallic phase) characterizing a re-entrant I-M transition, is seen. Another kind of bosonic MIT is observed at an insulator-superconductor transition [19] with intervention of a possible Bose-metallic intermediate phase in some cases: this has been called the "Bose metal", composed of pre-formed but phase-incoherent cooper pairs [20]. In this regard, Bismuth possesses the rare possibility of having Bose metallic phase [17] following the observation Das-Doniach scaling [21]. A distinct interpretation for the MIT at higher T is that it is driven by gap opening due to exciton formation and possible condensation, at least in Bi. The relation between these scenarios, if any, remains unclear at present and calls out for understanding.

Another key characteristic of these materials is giant magneto-resistance (GMR) of the order of ~$10^4$ to $10^6$ (MR%) [8-18]. This is in stark contrast to the generic observation of low MR as exhibited by non-magnetic materials and metals. To explain such GMR, which is independent of the material's magnetic properties, several mechanisms have been proposed: these include topological protection from backscattering in Dirac and Weyl semimetals (SMs) [10], classical multi-band model [12], and two band model with electron-hole compensation mechanism [8]. In multiband systems, Kohler scaling [22-25] provides information about the scattering processes involved in the transport of charge carriers and their temperature dependence: this can provide valuable insight into conditions where Landau Fermi liquidity, or it's breakdown, dominates physical responses. In materials close to a MIT, proximity to correlation induced local magnetic moments is expected to qualitatively modify this scaling, as seen, for example, in the normal state of cuprates and Fe arsenides. Finally, it is of interest that most of these materials exhibiting field driven MIT with GMR also often show Shubnikov de Haas (SdH) oscillations below the low T re-entrant IMT. The SdH oscillations at low temperatures (<15 K) are associated with an emergent 2DEG or surface conduction behaviour of the charge carriers [26]. In related contexts, it could also be that the observed upturn and plateau region at the re-entrant IMT at low T maybe associated with a scenario of a destruction of a possible topological surface state in the presence of magnetic field in absence of time-reversal symmetry [13]. An "organizing principle" that aims to rationalize the set of observations gleaned from various experiments is obviously of interest but remains elusive.

The pnictide elements As and Bi are well-established nearly compensated semimetals that exhibit large magnetoresistance under applied magnetic field [27,15]. Although previous studies have reported field-induced insulating behavior in Bi and discussed possible excitonic or superconducting correlations, a systematic examination of (i) re-entrant metal–insulator–metal transitions, (ii) the validity of Kohler scaling across distinct field regimes, and (iii) the applicability of two-parameter Das–Doniach (DD) scaling in the high-field regime remains lacking. In particular, the breakdown of Kohler scaling, which signals departure from semiclassical multiband transport, has not been quantitatively analyzed in gray arsenic, nor has its connection to DD-type scaling been explored.

In this article, we investigate the MIT transition driven by transverse magnetic field in pnictide elements i.e., gray Arsenic and Bismuth in their rhombohedral phases. Interestingly, Bismuth shows signatures of "double MIT" phase transition (a modulation in plateau region), which is rarely observed in such semimetals [10, 28]. The nature of scattering mechanism(s) followed by charge carriers in these elements is studied by analysis of Kohler's rule and (modified) Kohler's scaling whose unusual behaviour in high field regime indicates correlation effects in both Bi and As. The observed violation of Kohler scaling at higher fields motivates the application of DD two-parameter scaling, revealing a high-field scaling regime beyond classical compensation physics. To attempt a unifying rationalization of these features, we analyse the data within a theoretical picture based on a model of an unconventional, exciton fluctuation-mediated superconductivity in elemental Bi [29] proposed earlier by one of us.

## Experimental

The single crystalline sample of rhombohedral Bismuth and Arsenic is prepared by solid state reaction route following [29, 16]. The single crystal XRD of the grown crystal is shown in the supplement Fig. S(a), which aligns well with the previous report [26] and a recent study [30]. The gray Arsenic samples are the ones being reported in ref. 16. Both Bi and As have their crystal growth direction along c-axis. The transport measurements with transverse magnetic field (B||c-axis) were performed on Physical Property Measurement System (PPMS) by mounting the single crystalline sample on a PPMS DC puck with current flowing in ab-plane. The electrical contacts with copper wire were made following the four-probe geometry using silver EPOXY-H20E 10Z.

## Results and Discussion

Fig. 1(a) and (b) illustrates the variation in resistivity $\rho$ of the rhombohedral Bi and gray Arsenic single crystal with temperature (T) at zero field. The $\rho$ vs T plot for the As crystal resembles of a metal falling in Fermi liquid regime of electron-electron interaction at low temperatures, and at higher temperature the linear behaviour indicates the presence of electron-phonon and/or electron-electron scattering. More precisely, the curve behavior is explained by the equation below

$$\rho_{xx}(T) = \rho_{xx}(0) + \beta e^{\frac{-\theta}{T}} + \gamma T^2 \quad (1)$$

following P. Mal et al [31]. Here, $\rho_{xx}(0)$is the residual resistivity, the quadratic term represents electron-electron (*e-e*) interaction and the exponential term accounts for the electron-phonon (*e-ph*) interaction where $\theta$ being the Debye temperature. The $\rho(T)$ curve is fitted with the above equation as shown in Fig. 1(b) and the resulting fitted parameters are $\rho_{xx}(0) = 9.9 \times 10^{-8}$ mΩcm, $\beta = 2.25 \times 10^{-6}$ mΩcm, $\theta = 127.74$K and $\gamma = 2.97 \times 10^{-11}$ mΩcmK$^{-2}$. In case of Bismuth crystal, similar metallic behaviour is observed in $\rho - T$ plot, however a small hump is observed around 25 K-75 K. The insets of Fig. 1(a) and 1(b) shows the upturn in resistivity for both Bismuth and As crystals, as we decrease the temperature under applied transverse magnetic field. This upturn indicates the transition towards semiconducting behaviour, which is modelled by Arrhenius relation $\rho(T) \propto e^{\frac{E_g}{k_B T}}$ (where $E_g$ is the activation energy and $k_B$ is the Boltzmannn constant). Fitting the above equation to the observed upturns gives the activation energy for Bismuth as 24.5, 44.4, 58.8, 67.1, 72.5, 74.2, 80.6 meV for 1, 3, 5, 7, 9, 11 and 13 Tesla. The same activation energies for the As crystal are 1.52 meV, 3.33 meV and 4.0 meV at 6, 10 and 14 Tesla, respectively [16]. Thus, Bismuth has higher activation energy compared to that of Arsenic under transverse field. Correspondingly, for Bi, the upturn (semiconducting behaviour) starts at high temperature (~280 K) even at 1 Tesla field: it increases as the field increases. However, the low activation energy for As is comparable to that of other compensated semimetals [13].

Further, the magneto-transport data are analysed by calculating the *T*-derivative ($\frac{dR}{dT}$) of resistance at different applied fields which is shown in Fig. 1(c) and (d) for Bismuth and Arsenic respectively. The temperature derivative becomes more negative as we go down to temperature $T_{slope-min}$ and the characteristic plateau region starts below $T_{max}$, where the slope becomes positive. This shows the semiconducting region is more prominent near $T_{slope-min}$. In case of Bismuth, both $T_{slope-min}$and $T_{max}$, increase with the applied transverse magnetic field and they appear to saturate for higher fields as shown in Fig.1(e). For Arsenic, these two characteristic temperatures do not change significantly and it shows single MIT-like transition like other semimetals [8-16]. Interestingly, Bismuth shows a modulation in resistivity below $T_{max}$, indicating signature of a second, re-entrant IMT. The change in resistivity is comparatively low for this second MIT in comparison to first one. Following above Arrhenius model, we estimate the energy scale of the second MIT transition, which is found to be 0.05meV at 13 Test and 0.01meV at 3Tesla. Below $T_{max}$, the double MIT transition appear only above 1 Tesla field, which agrees with the previous reports [17, 18]. The transverse magnetic field driven double MIT is rare in such semimetals. Such undulation in plateau region is rarely reported. In TaAs, such modulations below $T_{max}$, are observed, and double MIT are shown at 5 and 7 Tesla [10], but below 3 Tesla and above 7 Tesla, only a single MIT was found.

Since magneto-transport measurement exhibit field driven MIT, GMR is observed for both Bismuth and Arsenic crystals. Fig 2(a) and (b) shows the MR% at various temperatures for both crystals, which is large, of the order of $10^5$%. For Bismuth, in the high-T metal, the MR shows usual metallic parabolic behaviour. But at low T, MR shows linear behaviour. The non-saturating linear MR in pure samples may indicate the contribution of topological Dirac electrons [32], or of excitons. In stark contrast, the MR of As shows parabolic behaviour for all T. In both cases, below 10 K, small undulations are observed at high fields; these are attributed to SdH oscillations. In Bismuth, although the first MIT takes place at 60 K, the SdH effect becomes resolvable only below 10 K, coinciding with the onset of the re-entrant IMT. This indicates that the optimal conditions for resolving high-mobility quantum transport occur below 10 K. Literature suggests that the optimal temperature for observing these quantum oscillations in semimetals is generally below 20 K [8-18]. The observed transport suggests an interplay between bulk and surface contributions. While field-driven MIT is a bulk phenomenon, the resistance saturation at high fields indicates that current is likely shunted through high-mobility surface channels. Although the magnetic field breaks time-reversal symmetry, these surface states remain sufficiently robust to prevent a total insulating divergence, leading to the observed re-entrant metallic plateau.

This indicate a possible onset of true Fermi liquid behaviour of electrons at low T, though caution is needed because the magnitude of the resistivity at low T is significantly higher than that of usual metals ("bad" metal). The de Haas-van Alphen effect, a sister phenomenon to SdH, is theoretically shown to work in the intermediate electron-electron interactions (as long as the Landau quasiparticle concept holds) and in the intermediate electron-phonon coupling regime as long as the cyclotron and phonon frequencies are comparable [33]. To observe SdH at higher temperatures, a tuning of cyclotron frequency and phonon oscillations is needed. However, for SdH in electron-phonon regime, a full theoretical framework is yet to be developed to our knowledge. In Bi, preformed excitons will couple to symmetry-adapted phonon modes, and their combined effect on SdH oscillations is an interesting, hitherto un-investigated avenue for future study.

To further understand the MR behaviour, Kohler scaling rule has been applied to the MR data to determine the contributions of various scattering processes involved in the Bismuth and Arsenic crystal samples. Here, the observed MR shows linear as well as quadratic dependencies on applied field as discussed above. It is useful to check Kohler scaling to determine the nature of scattering processes in various temperature ranges. Kohler's rule states that MR can be expressed as a function of $H\tau$, where $\tau$ represents the scattering time, defined as the duration between scattering events and is inversely related to zero-field resistivity [22, 23]. Physically, an increase in $H$ increases the cyclotron frequency, causing electrons to sweep out larger orbital angles between collisions. Consequently, the MR is determined as the function of $H\tau$(or equivalently $H/\rho_0$). For materials adhering to Kohler's scaling law, the response of the magnetoresistance depends solely on this product. Therefore, even as $\tau$ changes with temperature, the MR curves should overlap when plotted against $H\tau$. This collapse of data onto a single curve implies that a single, temperature-dependent scattering process is dominant.

Here, Fig. 3(a) and (b) show the variation in MR % with respect to $H/\rho_0$, i.e., $H\tau$ at different temperatures for Bismuth and As crystals, respectively. Interestingly, the MR % curves merge into each other at temperatures 5 K to 50 K for Bismuth but at 2 K and 100 K they show partial deviation. However, for As, all MR% curves merge except at 2 K. Hence, we observe partial violation of Kohler scaling rule. It may signify a significant change in scattering process at 2 K for both crystals. For Bismuth, all MR% curves at low temperature are found to be nearly parallel, but small deviations are observable at low fields similar to those reported in [34]. Interestingly, all MR% curves are nearly parallel to each other even at low fields for As crystal. It indicates that the extended Kohler's rule [19, 35] can be applied to As crystals. For this, an additional multiplier term to $H/\rho_0$ is added. This is $1/n_T$, temperature-dependent carrier density. Thus, in the extended Kohler rule, the MR% is expressed as a function of . Fig. 4(a) and (b) shows MR% vs $H/(\rho_0 n_T)$ curves at various temperatures for Bi and As respectively. Here, all MR% curves are scaled to the MR% curve at 100 K, i.e., for 100 K, the value of $n_T$ is taken to be 1. The observed values of $n_T$ for 5K to 50K are found to be ~0.88, however at 2K the same is 1.05 for Bismuth. The $n_T$ for As is ~0.95 at all other temperatures indicating that a single scattering mechanism is involved, except at 2K, where it is 1.29, which is quite a high value. The variation of $n_T$ with temperature is summarized in table 1. A typical pseudo gap temperature behavior of the $n_T$ is given by [36] $n_T = n_0 + \alpha T e^{-\Delta/k_B T}$, where $\Delta$ is pseudo-gap. As we decrease temperature, $n_T$ decreases and saturates to $n_0$. Here we observe clear deviation from the above guiding equation for both elements as shown in Table 1. This odd behavior of MR at low temperature is marked by the failure of extended Kohler rule. This unusual behavior of MR% needs further investigation in the context of the double MIT in Bismuth. The $n_T$ value of >1 indicates higher charge carrier density ($n_{2D}$) at low temperatures like $MoS_2$ hetero-structure[37] where $n_{2D}$ is higher than Hall charge carrier ($n_{e/h}$) measurements. Notably, we observe a clear breakdown of Kohler scaling at high magnetic fields and low temperatures, signaling a departure from classical multiband transport. Unlike earlier reports such as Zhao et al. [15], our analysis links this breakdown directly to the emergence of a distinct, non-classical high-field phase. Specifically, $n_T$ anomalously rises above 1 at ~2 K. We attribute this deviation to the lower-T re-entrant insulator-metal transition. Microscopically, as the field exceeds $B_c^{MIT}$, the field-induced excitonic condensate undergoes quantum melting (discussed in theoretical section). This melting releases localized carriers back into the conduction channel, driving both the sudden surge in carrier density and the concurrent failure of semiclassical scaling.

The higher-T MIT can be possibly understood in terms of a preformed exciton-induced gap opening. But the re-entrant IMT is more subtle in this scenario. We now present data for the possible signature of the Bose metal phase in these semimetals under transverse magnetic field near this IMT.

The Bose Metal (BM) phase is an intermediate metallic state in a superconductor insulator transition (SIT), which is usually a single quantum critical point (QCP) with one quantum of sheet resistance [19, 38] at $T = 0$. Under certain conditions, SIT is a two-step phase transition: a superconductor to BM phase followed by BM to Bose insulator phase generally observed in dirty (disordered) superconductor thin films [20] at low but finite temperature. The BM phase is characterized by formation of gapless but non-superfluid (i.e, phase-incoherent) cooper pairs leading to a Bose metallic phase. Das and Doniach showed the existence of BM phase at $T = 0$ emphasizing the role of two relevant parameters, viz. phase order associated with superconductivity, and charge, associated with the insulator [39]. Extending their work in a field driven scenario [21], they propose a two-parameter scaling rule

$$\frac{RT^{1+\frac{2}{Z}}}{\delta^{2\beta}} = f\left(\frac{\delta}{T^{1/z\upsilon}}\right) \tag{2}$$

where $\beta = \frac{\upsilon(z+2)}{2}$ and $\delta = B - B_{cr}$ for BM criticality. It has been shown that this works well at low temperatures. In the work by Kopelevich et al [17], three pairs of scaling exponents (i) $z = 1, \upsilon = 2/3$, (ii) $z = 1, \upsilon = 4/3$ and (iii) $z = 1, \upsilon = 2$ are discussed. Out of three class of critical exponents, the first two exponents scaling (i) $z = 1, \upsilon = 2/3$, (ii) $z = 1, \upsilon = 4/3$ have been reported to work well for $\delta \ll 1$ i.e. in the close vicinity of $B_{cr1}$ [20,39,40]. To perform the DD scaling, the $B_{cr1}$ is determined from magneto-transport data for Arsenic. The RT data at various fields as shown in inset of Fig. 5 is rescaled with $R/R_{35K}$ as shown in supplement Fig. S(b). The extracted value of $B_{cr1}$ for Arsenic is 1.5 Tesla. The scaling for Arsenic in close vicinity of $B_{cr}$ is shown in Fig. 5 with exponent $\upsilon = 4/3$, which works well. The same is also reported for Mo-Ge films [20, 21] and the Graphite samples [17]. The DD scaling in far field regime fails for the exponent, which we will also see for Bismuth and Arsenic. But the scaling is recovered for higher exponents. It is important to distinguish our scaling analysis from that of Ref. [17]. While earlier work applied the Das–Doniach framework to the low-field regime of Bi, our results demonstrate a scaling collapse in the high-field re-entrant regime where Kohler scaling fails. This extension of the Das–Doniach framework suggests that the Bose metal-like correlations are not limited to the initial insulating transition but govern the transport evolution across the entire high-field phase diagram.

Now we present the DD scaling for both elements in far field regime as shown in Fig. 6 and 7. For the scaling of Bismuth magneto-transport data, the $B_{cr1}$ (~0.035T) is taken from ref 17. Since, the data shown in the inset Fig. 1(a) and (b) are far away from $B_{cr1}$, they show significant deviation from the scaling as expected. For the large $\delta$, the scaling with critical exponents $z = 1, \upsilon = 2$ as suggested in ref. [19] shows significant improvement at low temperatures as shown in Fig. 6(c) and 7(c). This choice provides the best fit and, interestingly, coincides with the choice necessary to fit the B-dependent gap at the higher T MIT (see below). Even with the higher critical exponents as shown in Fig. 6(d) and 7(d), the scaling is almost equivalent. This shows the possible existence of BM phase at low temperatures for both Bismuth and Arsenic as they follow Das-Doniach type scaling. This is very unusual, since the BM phase is usually observed as a transient intermediate state at a SIT, but Bismuth is not a superconductor till 0.53 mK [41]. For Arsenic, no superconductivity is reported yet. However, if superconductivity is of the intermediate-to-strong coupling type, preformed, fluctuating cooper pairs will generically exist up to ``high'' T. This is a strong constraint on theoretical proposals.

**Theoretical Discussion**

It is very intriguing that the higher-*T* MIT and the lower-*T*, re-entrant IMT are seemingly described by distinct theoretical scenarios. Specifically, the higher-T MIT, occurring at $T_{MI}^{(1)}$, scales as $T_{MI}^{(1)}(B) \simeq \sqrt{B}$. This is consistent with the scenario of an insulating gap opening due to exciton formation [42]. On the other hand, observation of Das-Doniach scaling proximate to the lower-$T$ (at $T_{IM}^{(2)}$) suggests a Bose metal phase intervening between an incipient superconductor and a Bose insulator. Though neither SC nor (Bose) insulation obtains in our data (SC is only seen at much lower $T_c \simeq$ 0.5 mK in Bi, and not seen in As), our data suggest that associated critical fluctuations survive at the temperatures in our experiments. The central issue is: what is(are) the microscopic source(s) of these correlations in the distinct $T$ ranges close to $T_{MI}^{(1)}$ and $T_{IM}^{(2)}$ above? We now present a theoretical picture that addresses this issue qualitatively.

We begin with a specific first-principles correlated view, developed recently by one of us with collaborators [43]. In the realistic band structure of Bi, we have described the effects of a preformed, dynamically fluctuating excitonic liquid on normal state properties of Bi and on its eventual instability to a SC, in one picture. Good accord with a range of responses supports this multi-band view of Bi Metal. In this picture, both, excitonic liquid/solid and SC pairing tendencies arise from a single residual interaction. In the normal state of Bi, this effective interaction is generated by excitonic fluctuation exchange, in a way that is reminiscent of what happens in the well-studied Kondo-RKKY case in literature. The source of this effective interaction is the interband one-electron mixing in Bi, taken to second order *in the normal state*. At high $T$, it is irrelevant, and excitonic liquid correlations, coupled to inter-valley phonons, well describe the normal state transport responses. As $T$ is lowered, it becomes increasingly relevant. An external magnetic field also favors exciton formation at the expense of SC. As derived in [43], this effective interaction reads

$$H_{res} \simeq -\sum_{a,b,\sigma,\sigma'} t_{ab}^2 \chi_{ab} \left(a_{i,\sigma}^\dagger b_{j,\sigma} + h.c\right)\left(a_{i,\sigma'}^\dagger b_{j,\sigma'} + h.c\right) \tag{3}$$

It contains both excitonic ordering tendencies (via $a_{i,\sigma}^\dagger b_{j,\sigma} b_{i,\sigma'}^\dagger a_{j,\sigma'}$) and spin-singlet SC pairing instability (via $a_{i,\sigma}^\dagger a_{i,-\sigma}^\dagger b_{j,-\sigma} b_{i,\sigma} + h.c$). Crucially, in momentum space, the form-factor of the residual interaction is $\gamma(k)\gamma(k')$ where [43],

$$\gamma(k) = -\sqrt{3}\left(V''_{pp\sigma} - V''_{pp\pi}\right)\mathrm{si\,n}\left(\sqrt{3}\,k_x/2\right)\mathrm{si\,n}\left(k_y/2\right) \tag{4}$$

Note that it is even under spatial inversion and time-reversal ($k \rightarrow -k$ , but it breaks the discrete rotational symmetry. $H_{res}$ can be readily decoupled into two terms, in terms of two order parameters: the first $\Delta_e = \langle \gamma(k) a_{k,\sigma}^\dagger b_{k,\sigma} \rangle$, represents an excitonic condensate whenever finite, while the second, $\Delta_p^{(1)} = \langle a_{k,\sigma}^\dagger a_{-k,-\sigma}^\dagger \rangle, \Delta_p^{(2)} = \langle \gamma(k) a_{k,\sigma}^\dagger b_{-k,-\sigma}^\dagger \rangle$, describes a multi-band SC with intra- and inter-band Cooper pairing.

Since both these orders arise from the *same* single $H_{res}$, i.e, from the same set of bands, they are competing tendencies. Formally, this is seen by noticing that the usual *s*-wave pair operator, defined by $\frac{1}{\sqrt{2}}\sum_{k,\sigma}\sigma a^{\dagger}_{k,\sigma}a^{\dagger}_{-k,-\sigma}$, rotates $\Delta_e$ into $\Delta_p^{(2)}$. The critical region associated with emergence of these competing orders thus possesses an enhanced symmetry (multi-band systems). Application of an external field kills SC and enhances the competing excitonic order. As mentioned, $\Delta_e(k)$ breaks discrete rotational symmetry of $H$, so one expects an instability to an excitonic semiconductor with electronic nematicity. Once this happens, the $T$-dependence of the field-induced gap will follow the $T_{ex}(B) \simeq \Delta_e(B) \simeq B^{1/2}$ law [42], in full accord with our data (see Fig. (1)e, red curve), supporting an interpretation of the higher-$T$ MIT as one driven by exciton condensation. The energy scale of this higher-$T$ estimated by Arrhenius model ~ 25-80 meV (extracted from the experimental data in inset of Fig. 1(a)) is also comparable to excitonic energy scale in Bismuth [43], supporting excitonic semiconductor behavior. Using the same model for accompanying low-temperature re-entrant MIT, the energy scale is found to be 50 $\mu$eV, which is of the order (~100 $\mu$eV) of the observed electronic nematicity at the surface of Bismuth. This quantitative hierarchy suggests a single physical picture: the magnetic field initially stabilizes a robust excitonic gap (~25-80 meV), which subsequently undergoes quantum melting at higher fields. The resulting state is an incipient Bose liquid governed by ultra-low energy fluctuations ($\sim$ 50 μeV) and parallel surface conduction, preventing a total insulating divergence. Though the explicit exotic possibility of accompanying electronic nematicity remains to be shown in this regime, we point out that some support for this comes from the observation of a nematic quantum Hall liquid on the Bi-[111] surface [44].

However, it is also known that increasing $B$ beyond $B_c^{MIT}$ at which the higher-$T$ MIT occurs increases the carrier density. This weakens the tendency to exciton condensation, but one expects short-range, dynamical excitonic correlations to survive. The lower-T IMT can then be ascribed to the quantum melting transition of excitons due to this increased carrier density. This could also provide a possible hint of the increase in the T-dependent carrier number, n(T), at ~2 K in Bi. However, excitonic liquid-like correlations must survive across the IMT. Given the enhanced symmetry discussed above, one now also expects short-ranged, dynamical and "preformed" Cooper pair correlations to survive in the same regime. We propose that this is the origin of the "Bose metal" correlations that underpin the Das-Doniach scaling we observe. We believe that this feature is visible in semi-metals like Bi because of it's tiny Fermi pockets, and it should not occur in systems with higher carrier densities (larger Fermi surfaces). This principle ties both the exciton-driven MIT and Bose metal features to the mutually competing nature of incipient excitonic insulating and superconducting orders. Both are encoded in the very structure of the underlying residual interaction. It differs qualitatively from an earlier scenario [17], where the Das-Doniach scaling is associated with possible occurrence of "preformed" pairs in surface grains in Bi. Crucially, the application of the Bose metal (BM) framework here does not require a proximate transition to a global superconducting state, which occurs only at ~0.53 mK in Bi. Instead, the BM signature arises from the enhanced symmetry of the critical region dictated by $H_{res}$, which allows short-ranged, dynamically fluctuating preformed Cooper pairs and excitonic correlations to survive as a phase-disordered liquid at much higher temperatures. The Das-Doniach scaling effectively captures the dynamics of these incipient, competing bosonic fluctuations before phase coherence is established.

Within this proposal, $T_{ex}(B) \simeq \Delta_e(B) \simeq (B - B_c^{MIT})^{1/2}$, consistent with $z = 1$ and $\nu = 2$. Remarkably, we find that the Das-Doniach scaling that yields the best accord with data also occurs for $z = 1, \nu = 2$. In our opinion, this presents difficulties with an interpretation based on pairs in (randomly occurring) surface grains in Bi since, in that case, one might have expected either classical ($\nu = 4/3$) or quantum ($\nu = 8/3$) percolation to be relevant in scaling responses. We find that these choices do not result in acceptable scaling features, as discussed before. Furthermore, this state is truly anomalous, as it is known from earlier work [31] that the reduced magneto-resistance (MR) varies as $B^{1.3}$ (neither as $B$ nor as $B^2$), qualitatively similar to observations in cuprates and organics. Thus, this critical state is not likely to be a usual (Landau) Fermi liquid either. The relevance of preformed pairing and competing orders is extensively discussed in these cases [45], [46] but, to our best knowledge, has never been discussed in the context of elemental systems like Bi. Our interpretation of data in this work suggests that these ideas may also be relevant to field-induced quantum phase transitions in Bi.

Finally, observation of SdH-like oscillations at very low $T$ suggests a 2DEG formation in Bi. In our opinion, better understanding of this feature in a system with excitonic and pair fluctuation tendencies are non-trivial and requires more theoretical work.

**Conclusion**

In summary, in the present work magnetic field driven MIT in Bismuth and Arsenic single crystals have been studied. Arsenic crystal shows single MIT like other compensated semimetals, but Bismuth demonstrates signature of "double" MIT. Bi also shows SdH oscillations, pointing to a 2DEG formation under transverse magnetic field below 20K. The applicability of Kohler's scaling rule along with that of its extended version suggests that a single scattering mechanism is present in the temperature range up to $T_{max}$. At 2K, deviations from Kohler's rule imply an odd behaviour of MR suggestive of changes in charge carrier density. The microscopic origin of this last point is unclear, and more studies of the validity of Kohler's scaling in our theoretical model are necessary. A qualitative analysis based on a specific model of excitonic fluctuation-mediated superconductivity rationalizes both, the "higher T" MIT driven by exciton formation as well as the re-entrant I-M transition in terms of exciton melting and incipient Bose liquid in one single picture. Our study points to the exciting role of such exotic effects in shaping the physical responses of elemental pnictides.

**Supplementary Material**

Single crystal XRD characterization of the grown single crystals of Bismuth and scaled resistance plot of magneto-transport measurement of Arsenic to determine the critical field for metal to insulator transition.

**Acknowledgment**

The authors would like to thank the Director of the National Physical Laboratory, New Delhi for his keen interest. M. S.Laad thanks the DAE, Dovt of India, for support. N. K. Karn would like to thank CSIR, India for the research fellowship and Academy of Scientific and Industrial Research, Ghaziabad for Ph.D. registration. This work is supported by in-house project numbers OLP 240832 and OLP 240232.

**CRediT authorship contribution statement**

**N.K. Karn:** Writing – original draft, Methodology, Investigation, Data curation, Formal analysis, Writing – review & editing. **Mukul S. Laad:** Writing – original draft, Theoretical Analysis, Supervision, Methodology, Conceptualization, Writing – review & editing. **V.P.S. Awana:** Writing

– review & editing, Supervision, Project administration, Methodology, Conceptualization.

**Data Availability Statement:**

All the data associated with the MS will be made available at reasonable request.

**Conflict of Interest Statement and Declaration:**

The authors have no conflict of interest. The authors declare that they have no known competing financial interests or personal relationships that could have appeared to influence the work reported in this paper.

**Supplementary information:** Not Applicable

**Ethical approval:** Not Applicable

**Table1:** Relative charge carrier density ($n_T$) extracted from extended Kohler scaling at different temperatures

| T(K) | $n_T$ (As) | $n_T$ (Bi) |
|---|---|---|
| 2 | 1.29 | 1.1 |
| 10 | 0.94 | 0.9 |
| 20 | 0.95 | 0.9 |
| 50 | 0.98 | 0.9 |
| 100 | 1 | 1 |

**References:**

1. R. Zhang, Q. S. Fu, C. Y. Yin, C. L. Li, X. H. Chen, G. Y. Qian, C. L. Lu, S. L. Yuan, X. J. Zhao and H. Z. Tao, Understanding of metal-insulator transition in VO2 based on experimental and theoretical investigations of magnetic features, Scientific reports, **8**(1), 17093(2018).
2. B. Mallick, M. Palit, R. Jana, S. Das, A. Ghosh, J. Sunil, S. Maity, B. Das, T. Kundu, C. Narayana, and A. Datta, Pressure-induced insulator-to-metal transition in few-layer FePS3 at 1.5 GPa, Physical Review B, **109**(23), 235417(2024).
3. A. Giesekus, and L. M. Falicov, Theory of pressure-induced magnetic and metal-insulator transitions, Physical review B, **44**(19), 10449(1991).
4. S. Kettemann, Towards a comprehensive theory of metal–insulator transitions in doped semiconductors, Annals of Physics, **456**, 169306(2023).

5. R. Teruya, T. Sato, M. Yamashita, N. Hanasaki, A. Ueda and M. Matsuda, Reversible Insulator–Metal Transition by Chemical Doping and Dedoping of a Mott Insulator, Angewandte Chemie International Edition, **61**(34), e202206428(2022).
6. M. Imada, A. Fujimori and Y. Tokura, Metal-insulator transitions, Reviews of Modern Physics, **70**(4), 1039 (1998).
7. N. F. Mott, Metal-Insulator Transition, Reviews of Modern Physics, **40**(4), 677(1968).
8. M. N. Ali, J. Xiong, S. Flynn, J. Tao, Q. D. Gibson, L. M. Schoop, T. Liang, N. Haldolaarachchige, M. Hirschberger, N. P. Ong and R. J. Cava, Large, non-saturating magnetoresistance in WTe2, Nature, **514**(7521), 205 (2014).
9. O. Pavlosiuk, and D. Kaczorowski, Galvanomagnetic properties of the putative type-II Dirac semimetal PtTe2, Scientific Reports, **8**(1), 11297 (2018).
10. X. Huang, L. Zhao, Y. Long, P. Wang, D. Chen, Z. Yang, H. Liang, M. Xue, H. Weng, Z. Fang and X. Dai, Observation of the Chiral-Anomaly-Induced Negative Magnetoresistance in 3D Weyl Semimetal TaAs, Physical Review X, **5**(3), 031023(2015).
11. C. Shekhar, A. K. Nayak, Y. Sun, M. Schmidt, M. Nicklas, I. Leermakers, U. Zeitler, Y. Skourski, J. Wosnitza, Z. Liu, and Y. Chen, Extremely large magnetoresistance and ultrahigh mobility in the topological Weyl semimetal candidate NbP, Nature Physics, **11**(8), 645(2015).
12. S. Sun, Q. Wang, P. J. Guo, K. Liu and H. Lei, Large magnetoresistance in LaBi: origin of field-induced resistivity upturn and plateau in compensated semimetals, New Journal of Physics, **18**(8), 082002(2016).
13. F. F. Tafti, Q. D. Gibson, S. K. Kushwaha, N. Haldolaarachchige and R. J. Cava, Resistivity plateau and extreme magnetoresistance in LaSb, Nature Physics, **12**(3), 272 (2016).
14. P. Nowakowska, O. Pavlosiuk, P. Wiśniewski and D. Kaczorowski, Temperature-dependent Fermi surface probed by Shubnikov-de Haas oscillations in topological semimetal candidates DyBi and HoBi, Scientific Reports, **13**(1), 22776 (2023).
15. L. Zhao, Q. Xu, X. Wang, J. He, J. Li, H. Yang, Y. Long, D. Chen, H. Liang, C. Li and M. Xue, Magnetotransport properties in a compensated semimetal gray arsenic, Physical Review B, **95**(11), 115119 (2017).

16. N. K. Karn, K. Kumar, G. Awana, K. Yadav, S. Patnaik, and V. P. S. Awana, Magneto-transport and first principle study of strong topological insulator gray-arsenic, Mat. Res. Exp., **12**(3), 036301 (2025).
17. Y. Kopelevich, J. M. Pantoja, R. R. Da Silva, and S. Moehlecke, Universal magnetic-field-driven metal-insulator-metal transformations in graphite and bismuth, Phys. Rev. B, **73**(16), 165128 (2006).
18. N. K. Karn, Y. Kumar, G. Awana and V. P. S. Awana, Investigation of Angle-Dependent Shubnikov-de Haas Oscillations in Topological Insulator Bismuth, Physica Status Solidi (b), 2400077(2024).
19. P. Phillips, and D. Dalidovich, The Elusive Bose Metal Science, 302(5643), 243 (2003).
20. N. Mason and A. Kapitulnik, Dissipation Effects on the Superconductor-Insulator Transition in 2D Superconductors, Phys. Rev. Lett., **82**, 5341 (1999).
21. D. Das, and S. Doniach, Bose metal: Gauge-field fluctuations and scaling for field-tuned quantum phase transitions, Phys. Rev. B, **64**(13), 134511 (2001).
22. I. L. Spain and R. O. Dillon, Kohler's rule and other scaling relationships for the magneto-resistance of graphite, Carbon, **14**(1), 23 (1976).
23. A. Malinowski, V. L. Bezusyy, and P. Nowicki, Saturation of resistivity and Kohler's rule in Ni-doped La1.85Sr0.15CuO4 cuprate, Phys. Rev. B, **95**(1), 014521 (2017).
24. J. Xu, F. Han, T. T. Wang, L. R. Thoutam, S. E. Pate, M. Li, X. Zhang, Y. L. Wang, R. Fotovat, U. Welp and X. Zhou, Extended Kohler's Rule of Magnetoresistance, Physical Review X, **11**(4), 041029 (2021).
25. N. K. Karn, M. M. Sharma and V. P. S. Awana, Weak antilocalization and topological edge states in PdSn4, Journal of Applied Physics, **133**(17) (2023).
26. Z. Ren, A. A. Taskin, S. Sasaki, K. Segawa and Y. Ando, Large bulk resistivity and surface quantum oscillations in the topological insulator Bi2Te2Se, Physical Review B, **82**(24), 241306 (2010).
27. Zhu, Z., Fauqué, B., Behnia, K. and Fuseya, Y., Magnetoresistance and valley degree of freedom in bulk bismuth, Journal of Physics: Condensed Matter, **30**(31), 313001 (2018).
28. N. Kumar, N. K. Karn, P. Kushwaha and V. P. S. Awana, Growth, Characterization and Magneto-transport of single crystal Bismuth, Indian Journal of Engineering and Materials Sciences (IJEMS), **31**(3), 335 (2024).

29. Y. Kumar, P. Sharma, N. K. Karn and V. P. S. Awana, Shubnikov-de haas (sdh) oscillation in self-flux grown rhombohedral single-crystalline bismuth, J. Supercond. Nov. Magn., **36**(2), 389 (2023).
30. A. Sáenz-Hernández, S. Sangiao, C. Felser, C. Shekhar, J. A. Pardo, A. Ibarra, and J. M. De Teresa, Cryogenic Focused Ion Beam Milling to Investigate the Anisotropic Magneto-transport Properties of Bismuth Microcrystals. *Advanced Functional Materials*, e17475 (2025).
31. P. Mal, B. Das, A. Lakhani, G. Bera, G. R. Turpu, J. C. Wu, C. V. Tomy and P. Das, Unusual Conductance Fluctuations and Quantum Oscillation in Mesoscopic Topological Insulator PbBi4Te7, Sci. Rep., 9(1), 7018 (2019).
32. Y. Sun, T. Taen, T. Yamada, S. Pyon, T. Nishizaki, Z. Shi and T. Tamegai, Multiband effects and possible Dirac fermions in Fe1+$y$Te0.6Se0.4, Physical Review B, **89** (2014) 144512.
33. M. Fowler and R. E. Prange, Electron-Phonon renormalization effects in high magnetic fields, Physics Physique Fizika, **1**(6), 315 (1965).
34. J. M. DeStefano, E. Rosenberg, O. Peek, Y. Lee, Z. Liu, Q. Jiang, L. Ke and J. H. Chu, npj Quantum Materials, **8**(1), 65 (2023).
35. N. H. Jo, Y. Wu, L. L. Wang, P. P. Orth, S. S. Downing, S. Manni, D. Mou, D. D. Johnson, A. Kaminski, S. L. Bud'ko and P. C. Canfield, Extremely large magnetoresistance and Kohler's rule in PdSn4: A complete study of thermodynamic, transport, and band-structure properties, Physical Review B, **96**(16), 165145 (2017).
36. J. Xu, F. Han, T. T. Wang, L. R. Thoutam, S. E. Pate, M. Li, X. Zhang, Y. L. Wang, R. Fotovat, U. Welp, and X. Zhou, Extended Kohler's rule of magnetoresistance. *Physical Review X*, *11*(4), 041029 (2021).
37. X. Cui, G. H. Lee, Y. D. Kim, G. Arefe, P. Y. Huang, C. H. Lee, D. A. Chenet, X. Zhang, L. Wang, F. Ye and F. Pizzocchero, Multi-terminal transport measurements of MoS2 using a van der Waals heterostructure device platform, Nature nanotechnology, **10**(6), 534 (2015).

38. M. P. Fisher, G. Grinstein and S. M. Girvin, Presence of quantum diffusion in two dimensions: Universal resistance at the superconductor-insulator transition, Phys. Rev. Lett., **64**(5), 587 (1990).
39. D. Das, and S. Doniach, Existence of a Bose metal at $T$ =0, Phys. Rev. B, **60**(2), 1261 (1999).
40. S. Moehlecke, Y. Kopelevich and M. B. Maple, Local superconductivity and ferromagnetism interplay in graphite-sulfur composites. Brazilian Journal of Physics, **33**, 762 (2003).
41. O. Prakash, A. Kumar, A. Thamizhavel and S. Ramakrishnan S, Evidence for bulk superconductivity in pure bismuth single crystals at ambient pressure, Science, **355**, 52 (2017).
42. Khveshchenko, D.V., Magnetic-field-induced insulating behavior in highly oriented pyrolitic graphite. Phys. Rev. Lett., **87**(20), 206401 (2001).
43. S. Koley, M. S. Laad and A. Taraphder, Dramatically enhanced superconductivity in elemental bismuth from excitonic fluctuation exchange, Scientific Reports, **7**, 10993 (2017).
44. B. E. Feldman, M. T. Randeria, A. Gyenis, F. Wu, H. Ji, R. J. Cava, A. H. MacDonald and A. Yazdani, Observation of a nematic quantum Hall liquid on the surface of bismuth. Science, **354**(6310), 316 (2016).
45. Yayu Wang, Lu Li and N. P. Ong, Nernst effect in high- superconductors, Phys. Rev. B, **73**, 024510 (2006).
46. K. Fujita, M. H. Hamidian, S. D. Edkins, C. K. Kim, Y. Kohsaka, M. Azuma, M. Takano, H. Takagi, H. Eisaki, S. I. Uchida and A. Allais, Direct phase-sensitive identification of ad-form factor density wave in underdoped cuprates. Proceedings of the National Academy of Sciences, **111**(30), E3026 (2014).

## Figures with Captions:

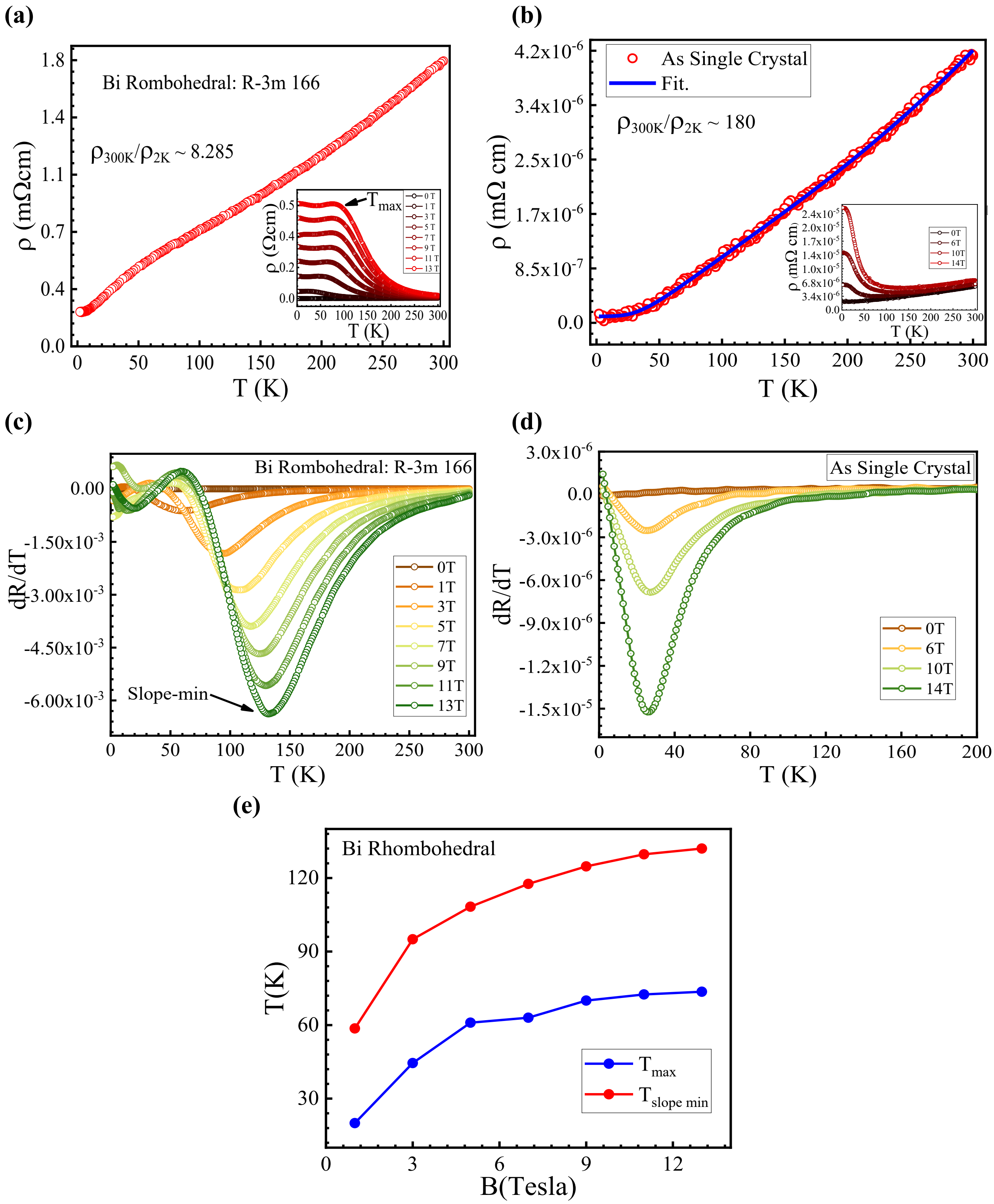

**Figure 1** Temperature dependent resistivity (ρ-T) plot at zero field and the inset shows ρ-T at different transverse field for **(a)** Bi crystal and **(b)** As crystal reproduced from ref. [16]. Temperature derivative of resistance at different transverse field for **(c)** Bi crystal and **(d)** As crystal. **(e)** Characteristic temperature variation with applied field for Bi crystal.

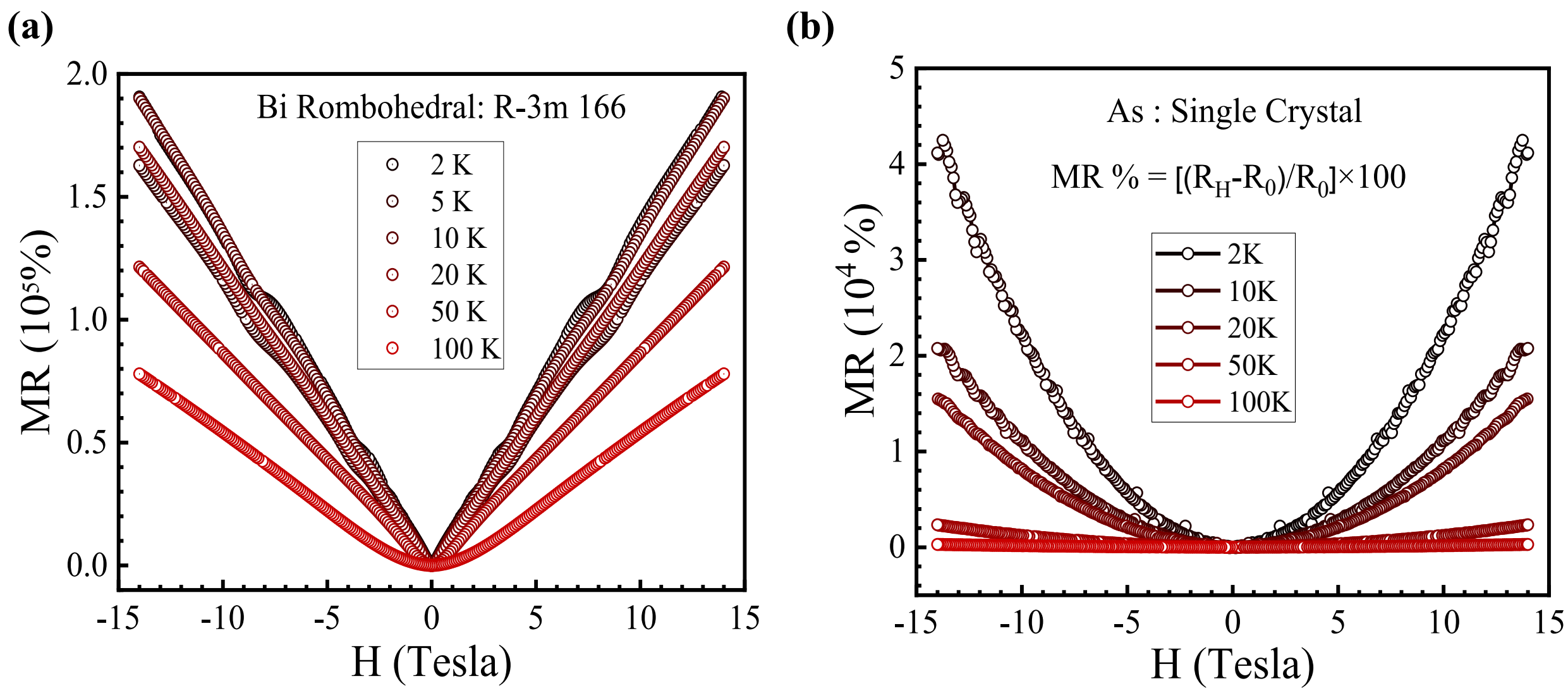

**Figure 2 (a)** and **(b)** shows the magnetoresistance of Bi and As crystals at different temperatures, respectively.

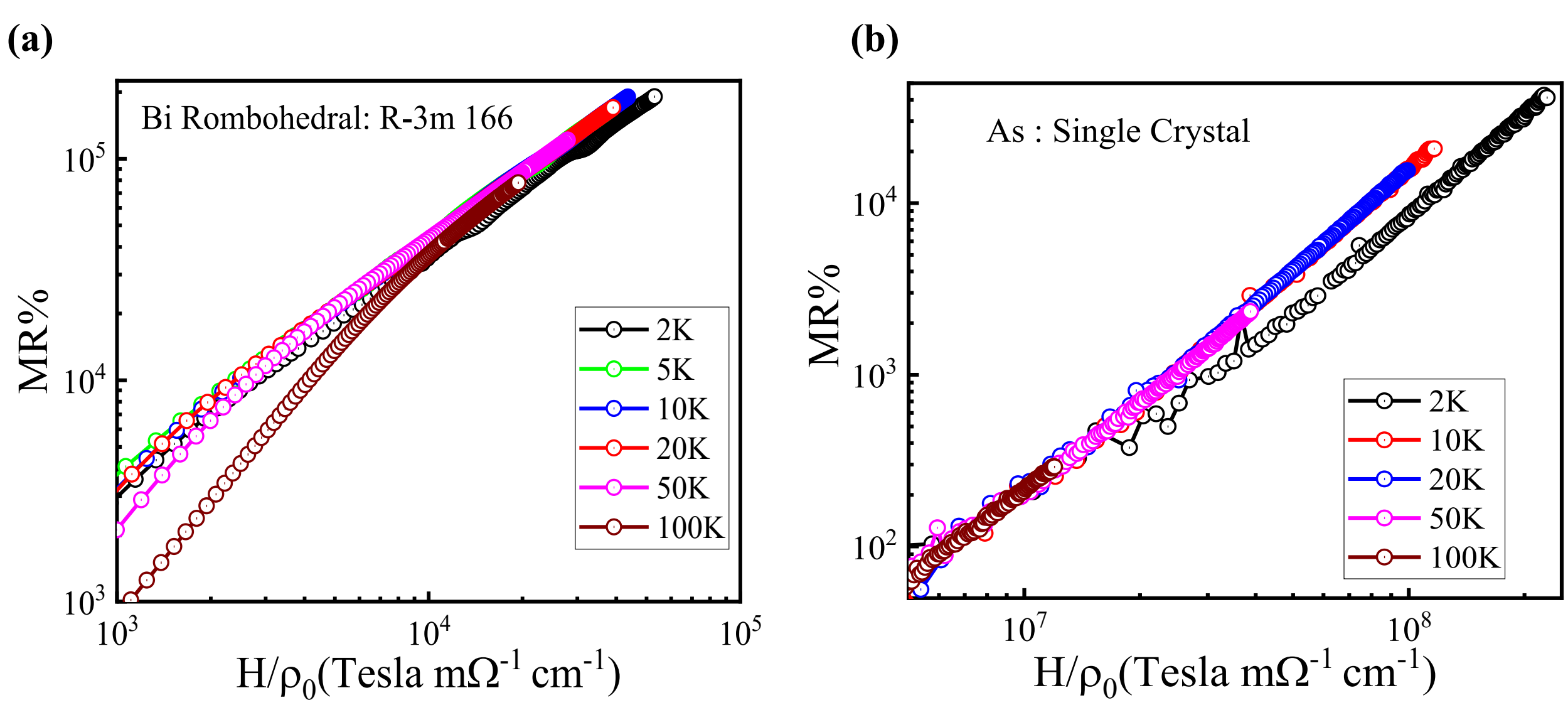

**Figure 3 (a)** and **(b)** shows the Kohler scaling (MR% vs $H/\rho_0$ plot) for Bi and As single crystals, respectively.

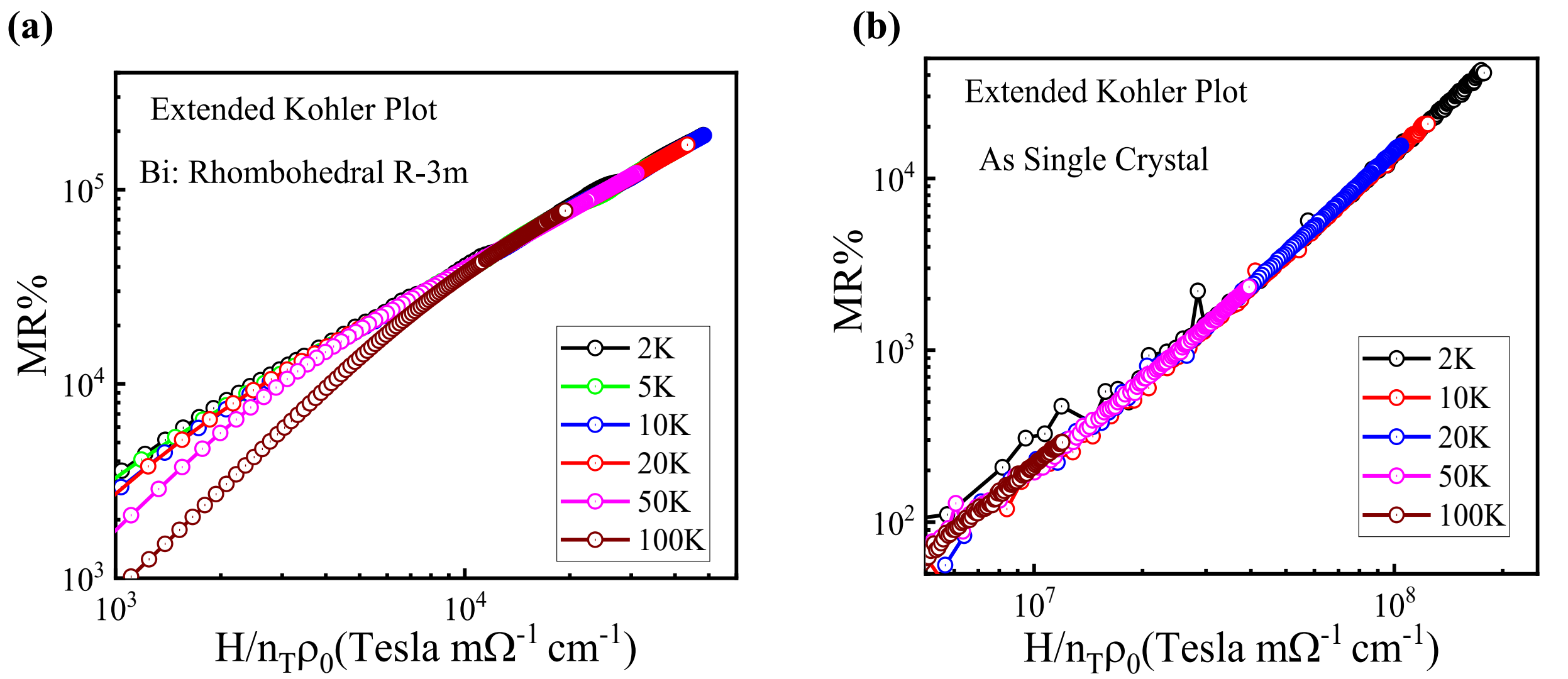

**Figure 4 (a)** and **(b)** shows the Extended Kohler scaling (MR% vs $H/n_T\rho_0$ plot) of magnetoresistance for Bi and As crystal, respectively.

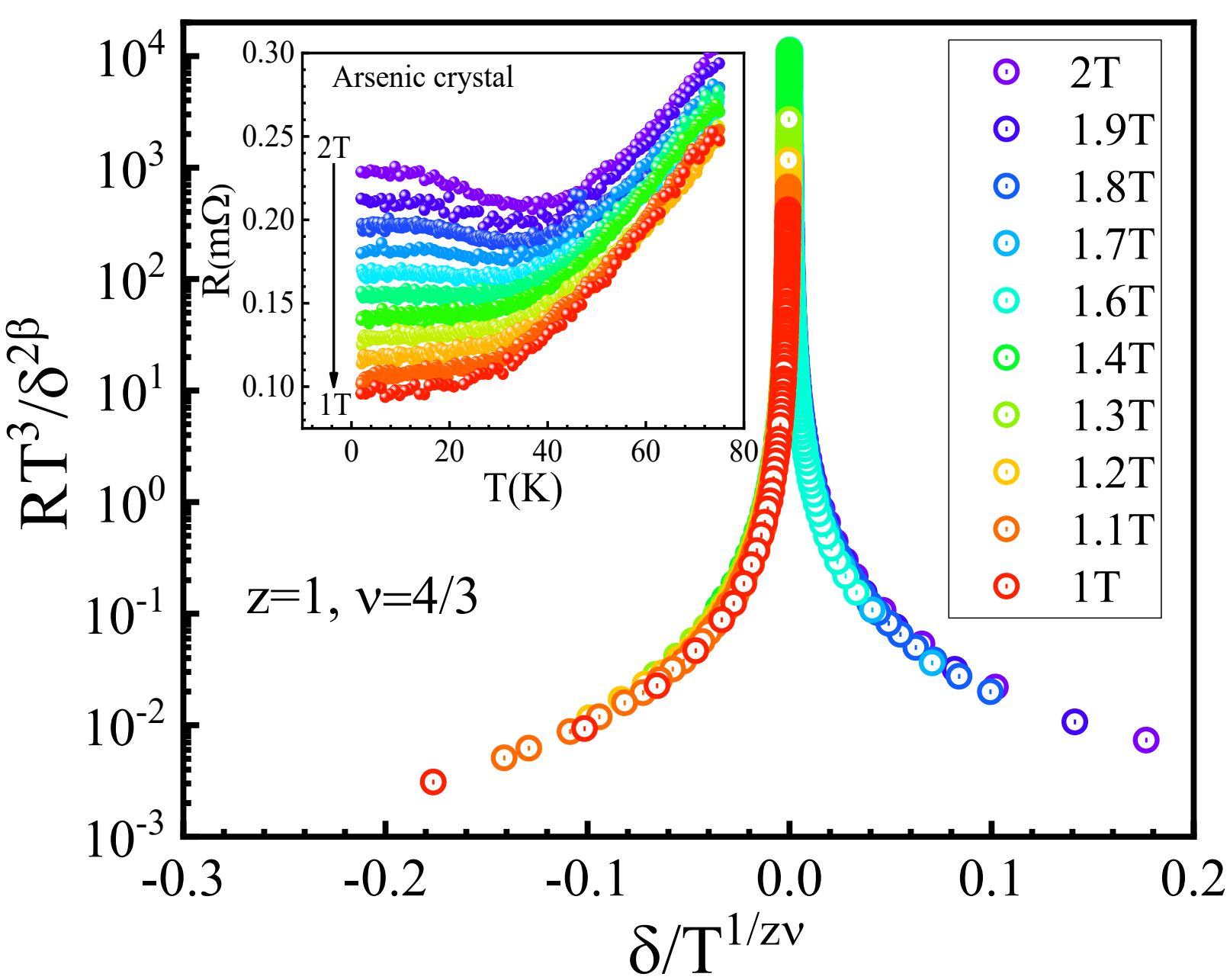

**Figure 5** Bose metal-insulator transition analysis of magneto-transport data for Arsenic crystal in near critical field regime.

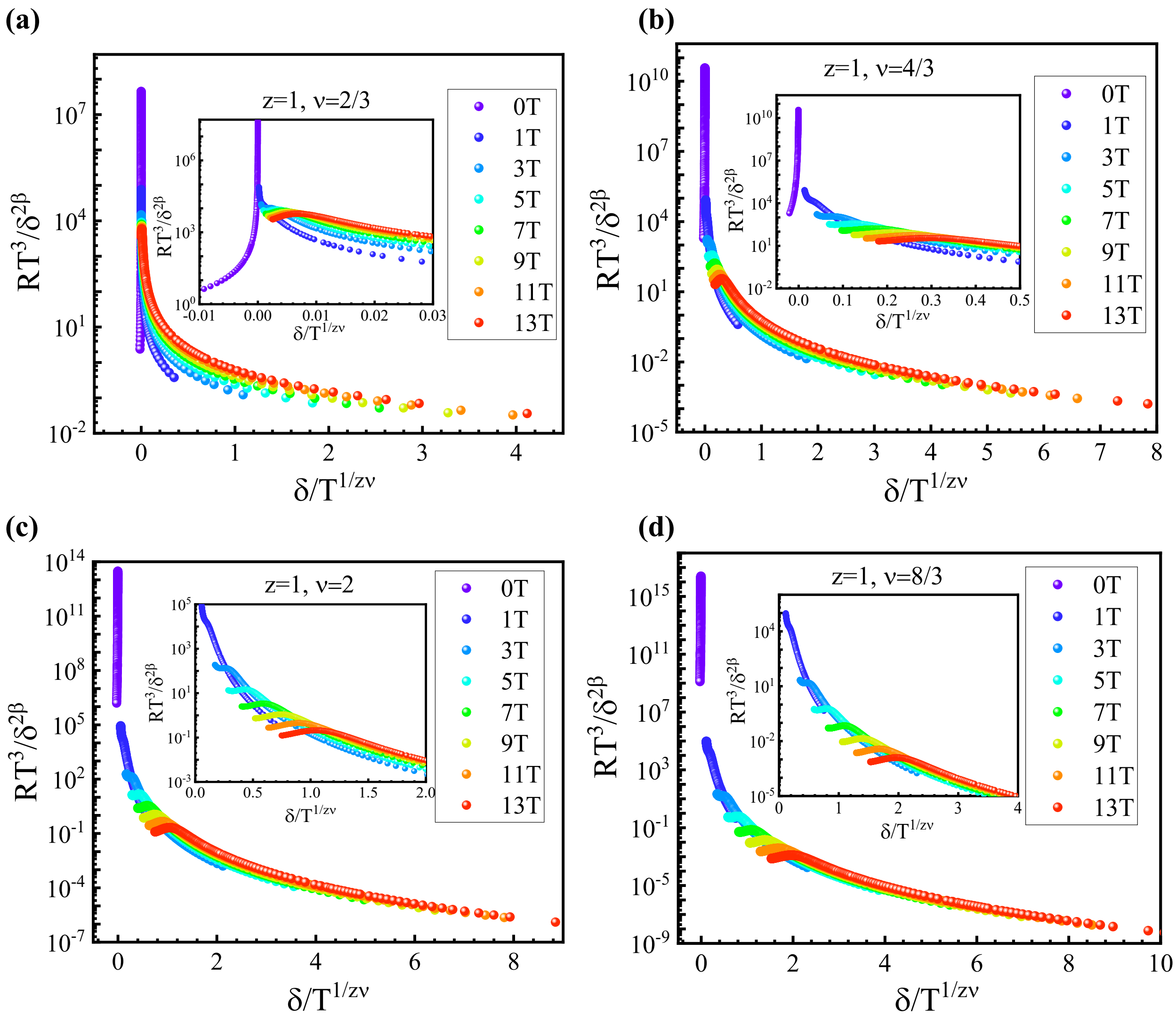

**Figure 6** Das and Doniach type scaling of ρ-T data for Bismuth (data used is from the inset of Fig. 1(a)) (a) z=1, ν=2/3 (b) z=1, ν=4/3 (c) z=1, ν=2 (d) z=1, ν=8/3

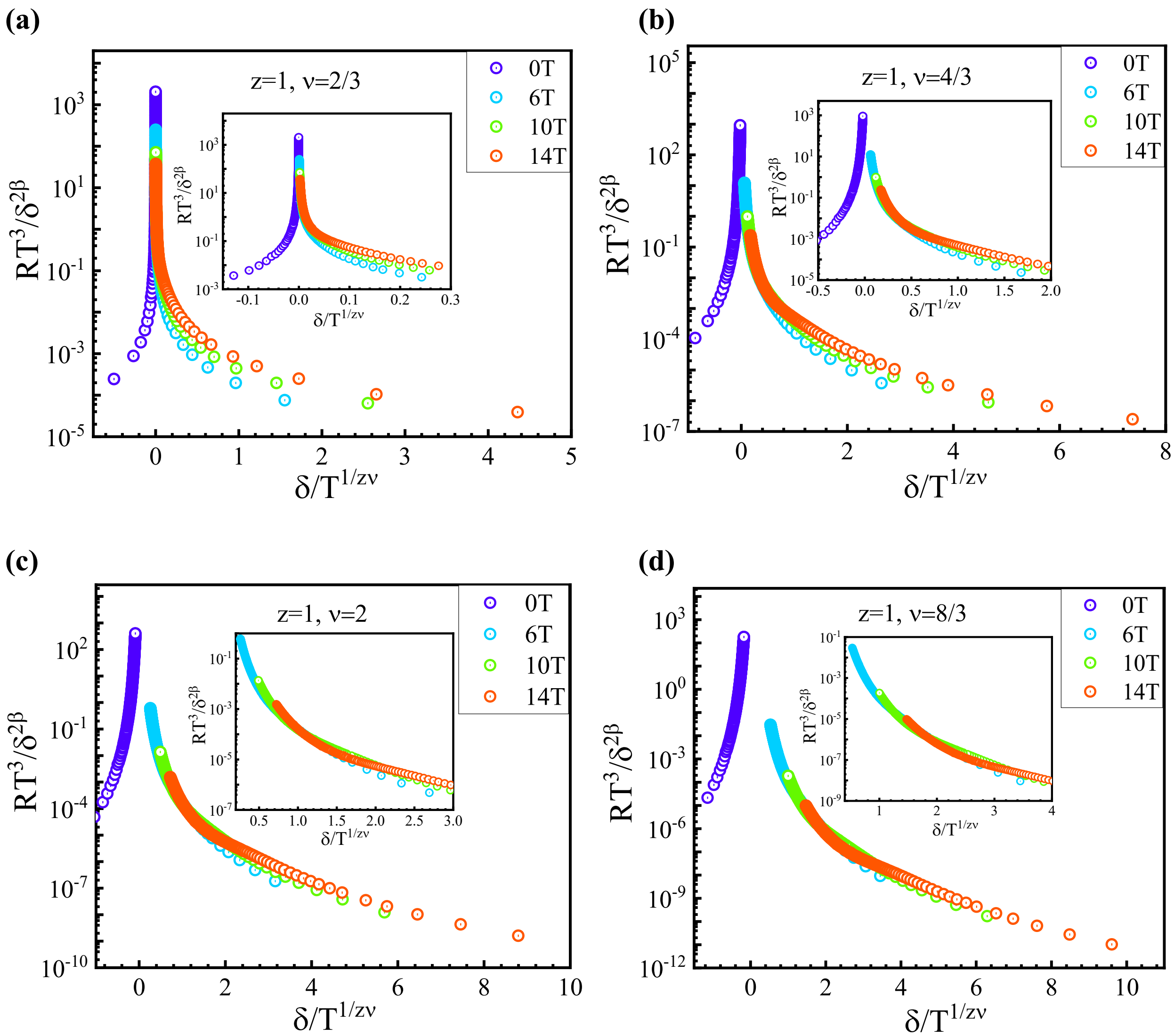

**Figure 7** Das and Doniach type scaling of ρ-T data for Arsenic (data used is from the inset of Fig. 1(b) of the MS) (a) z=1, ν=2/3 (b) z=1, ν=4/3 (c) z=1, ν=2 (d) z=1, ν=8/3